\title[RankBooster: Visual Analysis of Ranking Predictions]%
      {RankBooster: Visual Analysis of Ranking Predictions}

\author[Abishek Puri et al.]
{\parbox{\textwidth}{\centering
Abishek Puri$^{1}$\orcid{0000-0001-8945-6984},
Bon Kyung Ku$^{1}$\orcid{0000-0002-4534-8110},
Yong Wang$^{1}$\orcid{0000-0003-0177-1731},
Huamin Qu$^{1}$\orcid{0000-0002-3344-9694}
        }
\\
{\parbox{\textwidth}{\centering 
          $^1$ Hong Kong University of Science and Technology, Hong Kong
       }
}
}

\begin{document}


\maketitle
\begin{abstract}
Ranking is a natural and ubiquitous way to facilitate decision-making in various applications. However, different rankings are often used for the same set of entities, with each ranking method placing emphasis on different factors. These factors can also be multi-dimensional in nature, compounding the problem. This complexity can make it challenging for an entity which is being ranked to understand what they can do to improve their rankings, and to analyze the effect of changes in various factors to their overall rank. 
In this paper, we present RankBooster, a novel visual analytics system to help users conveniently investigate ranking predictions. We take university rankings as an example and focus on helping universities to better explore their rankings, where they can compare themselves to their rivals in key areas as well as overall. Novel visualizations are proposed to enable efficient analysis of rankings, including a \textit{Scenario Analysis View} to show a high-level summary of different ranking scenarios, a \textit{Relationship View} to visualize the influence of each attribute on different indicators and a \textit{Rival View} to compare the ranking of a university and those of its rivals. A case study demonstrates the usefulness and effectiveness of RankBooster in facilitating the visual analysis of ranking predictions and helping users better understand their current situation.


\end{abstract}  
\begin{figure*}
  \includegraphics[width=\textwidth,height=7cm]{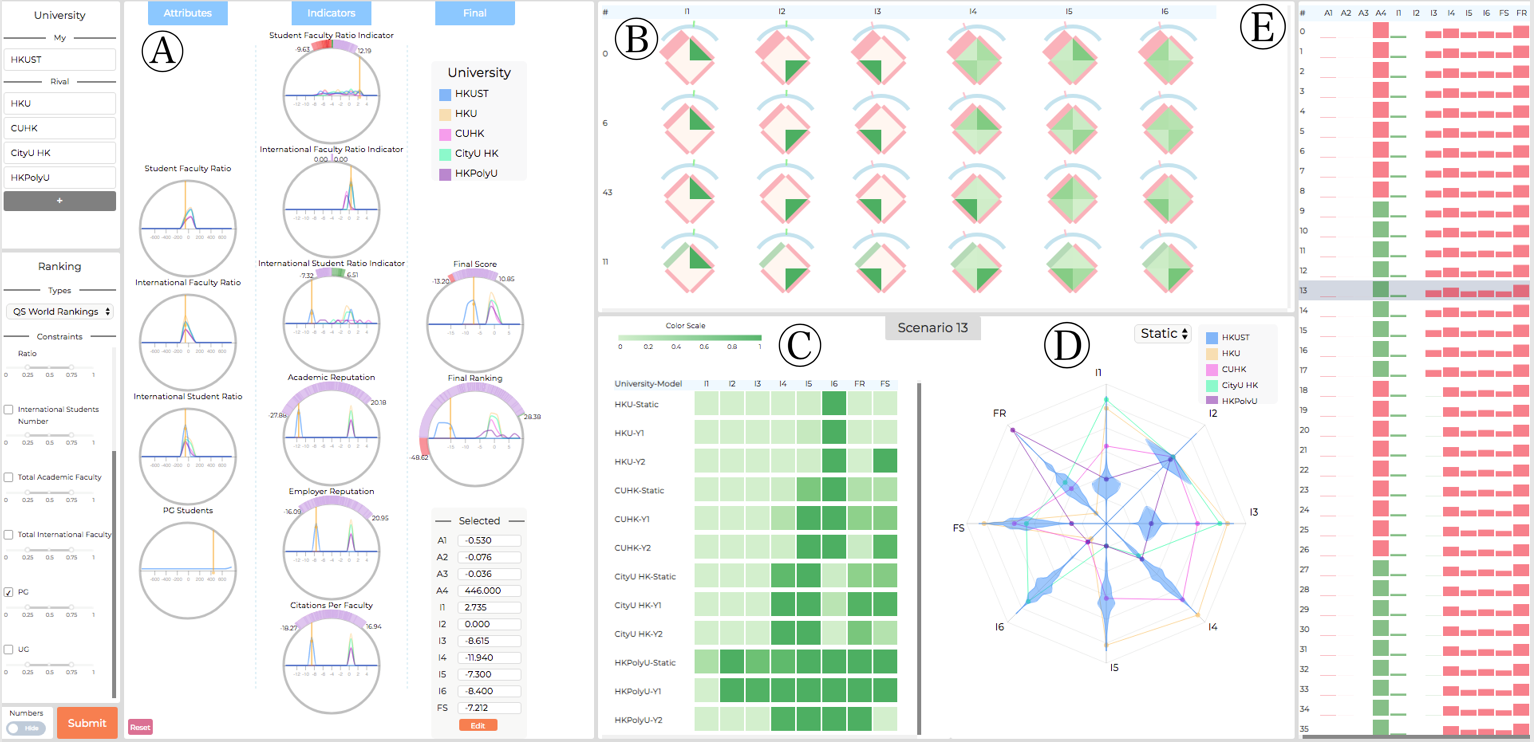}
   \caption{The user interface showing multiple rival universities, where Scenario 13 is selected for further analysis. (A) the Scenario Analysis View, (B) the Relationship View where 4 scenarios have been selected for comparison, (C) the Rival Heat Map View, (D) the Rival Radar Chart View, (E) the Scenario List View. }
   	\label{fig:teaser}
\end{figure*}
\section{Introduction}

Ranking is a natural and ubiquitous way to help people make decisions in various applications, for example, finding a good restaurant and hotel for a trip, choosing the best university, and buying a suitable car. Many studies have been conducted to help people easily interpret rankings and make better decisions through visualization. Few studies, however, have focused on helping the entities (or items) being ranked, which we call ``rankees''.

Due to the wide usage of ranking companies, the impact of a low ranking can be substantial. For example, universities with a lower ranking receive less applications and private  \cite{Rankings_Impact}. In this paper, we aim to tackle the following question \textit{how can we use visualization techniques to help universities efficiently interpret their rankings and analyze the effect of their actions on their ranking?} 

There are three reasons why this is a challenging question: multiple ranking systems and large number of entities, prediction uncertainties, and users' unfamiliarity with ranking prediction and visualization.

First, the sheer number of systems and methodologies for creating rankings creates complexity. Second, predictions are innately uncertain, and so the system must show the user the uncertainty. Third, many of the university administrators are not familiar with predictive analytics and visualization, requiring the system to be simple yet comprehensive.

Our research problem can be regarded as a Multi-Criteria-Decision-Making (MCDM) problem with the added complexity of uncertainty visualization. There have been many prior studies in the field of MCDM over the past decade \cite{weightlifter,mcdm_fisheries,mcdm_mograms,mcdm_top,krause2016interacting}, but they have generally focused on either high-level analysis of the tasks required to be solved \cite{high_level_MCDM} or aiding decision-making without any uncertainty involved in the variables {\cite{weightlifter}}. To the best of our knowledge, in the domain of ranking visualization, there has been no work attempting to aid rankees in their decision-making issue.

In this paper, we propose a visual analytics system, RankBooster, built in collaboration with university ranking experts. During this collaboration, we were able to create a set of tasks that could be generalized to other sets of entities who need to understand how to improve their ranking position in their ranking systems.

Our visual analytics system was used by our university to analyze their ranking position prior to their data submission. Our case study with the domain experts also shows the effectiveness of our approach.

Our major contributions can be summarized as follows:
\begin{itemize}
    \item A visual analytics system, RankBooster.
    \item A novel glyph visualization to show the relationship between the Indicators and the Attributes in ranking systems. 
    \item A case study with the domain experts highlighting the usability of the system.
\end{itemize}

\section{Related Work} 

As our work comes from the ranking visualization domain, we will list out key work in ranking visualization.

Part of the literature consists of works that aim to allow users to interact with the ranking system by modifying the weights or inputting their preferences. Works in this area include LineUp \cite{lineup}, Podium \cite{Podium}, and Rankit \cite{rankit}. However, they do not attempt to visualize any predictions of established rankings, focusing on existing rankings and making them more interactive and accessible. Other work has focused on looking at rankings over time, allowing users to observe trends and patterns. An example work in this area is RankExplorer \cite{rankexplorer}. While such works allow rankees to evaluate historical trends visually, they do not provide a way to evaluate scenarios in the future, or evaluate the effect of decisions made now. With regards to university rankings, which is the domain we will focus on in the paper, we could find only one other work \cite{dcpairs_uses_universities_ranking} that uses university rankings, but again their focus is on helping students make the decision, not the universities. Therefore, we claim that our perspective on this domain is a novel one.

\section{Background}
 
\subsection{Terminology}
 In this subsection, we explain key terminology used in the paper, through an example of a ranking system submission and result.
 
 First, raw data is collected on the rankees. These are the \textit{Attributes}. Then, the data is split into different groups and normalized across all rankees. These normalized groups are the \textit{Indicators}, and the normalized value for a given rankee in an Indicator is their \textit{Indicator Score}. Finally, all of a rankees Indicator Scores are inputted into a function which then gives them their \textit{Final Score}. These final scores are the basis of a rankees final rank.
 
 The question that is of interest to the rankee is \textit{What will be the Final Score given the values of the Attributes I am submitting?} To frame the question, we create a \textit{Scenario}, which consists of a set of Attributes and their values, the predicted Indicator Scores, and the predicted Final Score. Our system allows the user to evaluate scenarios individually or collectively.
 
 \subsection{Predictive Model Requirements}
 
 Our system assumes the user already has a model that can generate predictions of the Indicator Scores, along with some measure of uncertainty. It also assumes that the user already has a model that can generate predictions for other universities, along with some measure of uncertainty. 
 
 During our collaboration, an ensemble model which produced 100 predictions for each Indicator Score was used, with the range of the predictions providing us with the measure of uncertainty. 
 
 \subsection{Example Of A Scenario}
 
 As an example, we imagine a ranking system with 3 Attributes, coalesced into 2 Indicators, eventually creating a final score. In a given scenario, we have known values for the 3 Attributes. Then, we have predictions for the 2 Indicator Scores, as well as a prediction of the Final Score. These predictions are normalized to be between 1 and a 100 inclusive. For each prediction, we also have a measure of uncertainty. In our case, the uncertainty is the range of the 100 values produced by the model for each prediction. 



\subsection{Dataset Access}
Our dataset includes 5 years of data on over 400 universities. While the Indicator and Final Score are publicly available through the respective ranking systems websites, the Attribute values are not publicly available, requiring us to get the dataset from our collaborators.

\section{Requirements Analysis}

Based on a survey of ranking experts, we were able to identify 3 main requirements of the system. 
\begin{itemize}
    \item \textbf{R1. Explore Scenarios Individually} 
    The system should be able to track a scenario from the submitted data to the final score. It should also show the relevant uncertainties, and identify Attributes and Indicators that will have the most impact on their rankings.
    
    \item \textbf{R2. Explore Scenarios As A Group}
    The system should be able to compare Attributes, Indicators, Final Scores, and uncertainties between scenarios. Users should also be able to filter and constrain the set of scenarios, and visualize a summary.
    
    \item \textbf{R3. Compare with Rival Universities} 
    The system should be able to directly compare a selected university with its rivals, focusing on predicted differences given a particular scenario.
\end{itemize}


    
    

\section{Visual Design}
This section describes the usage of each view in our system, the design rationale for each view, and the interactions in each view.

\subsection{Scenario List View}
This view, shown in Fig.\ref{fig:teaser}(E), is a table, with each row of the table showing a summary of one scenario. This view partially satisfies \textbf{R2}, specifically providing a filterable and sort-able summary of all scenarios. It gives the user a simple interface by which to select particular scenarios for further analysis in the other views of the system. 

We show the change in the Attributes and the predicted change in Indicator Scores and Final Scores as bars. Here, the change is relative to the previous years values and scores respectively. The color represents whether the change is positive (green) or negative (red). 



\subsection{Scenario Analysis View}
This view, shown in Fig.\ref{fig:teaser}(A), consists of three columns of glyphs. The leftmost column has a glyph per Attribute, the middle column has a glyph per Indicator, and the rightmost column has a glyph each for the Final Score and the Final Rank. This view partially satisfies \textbf{R2}, showing a group summary of all the scenarios along with the uncertainty related to the predictions being made. It also partially satisfies \textbf{R3}, as multiple entities can be compared in this view.

Each glyph shows all the values for that object across all the selected scenarios, relative to the previous value of that object. For example, a glyph in the Attribute column would show all the values for a particular Attribute across all the scenarios, relative to the value of the Attribute in the previous year. This is encoded with a line chart, with the x-axis containing the relative values and the y-axis containing the frequency of appearance in the scenarios. 

We encode uncertainty by enhancing the ensemble bar from Chen et al. \cite{QU_uncertainty}. The enhancement consists of allowing multiple entities to be compared by superimposing each entities uncertainty on each other, whilst giving each entity a different color. The length of the bar along the glyph corresponds to the range of the uncertainty, with the lowest and highest relative values in the range of the prediction shown at the left and right of the bar respectively. In this encoding, negative values depict being below the previous years values. 

\subsection{Relationship View}
The view, shown in Fig.\ref{fig:teaser}(B), is a table of glyphs. Each row of the table represents a selected scenario, with each glyph in the row encoding information about the Attributes in the scenario and one Indicator in the scenario. Each column of the table represents one particular Indicator. This view partially satisfies \textbf{R1} and \textbf{R2}, specifically allowing the user to analyze the effects of the Attribute values of a scenario on the predicted Indicator Score and Final Score, whilst also allowing multiple scenarios to be compared with each other along the Indicator axis.

Any given glyph in this view is linked to a particular scenario and a particular Indicator. In Fig.\ref{fig:teaser}(B), the top-left glyph is linked to Scenario 0 and Indicator 1. The square shape of the glyph is due to the number of attributes; if there are more attributes, the shape would be a higher-sided polygon. The bar on the side of the glyph shows the value of a particular attribute in that scenario relative to the previous years value. The polygon is internally split equally amongst the attributes, with the shade signifying the relationship between that attribute and the indicator. The darker green the color of the section, the more the Indicator Score increases when the Attribute increases. The opposite is true if the color is a dark red. The curved bar on top of the glyph shows the predicted Indicator Score for that scenario with a line, and the maximum and minimum value of the Indicator Score is the total length of the bar. 

We determine the effect of an Attribute on the Indicator Score by perturbing the Attribute value and observing how that affects the models' Indicator Score prediction. The effect is normalized across all scenarios selected to be in the view.

\subsection{Rival View}
This view has a heat map, shown in Fig.\ref{fig:teaser}(C), and a radar chart, shown in Fig.\ref{fig:teaser}(D). This view satisfies \textbf{R3}, allowing the user to compare predictions for different universities. The heat map encodes the probability that the users university will have a higher Indicator or Final Score than a rival university, given a particular scenario. Our users had 3 different methods for predicting the Indicator or Final Score for the rival universities; each method is given a row in the heat map. 

The radar chart encodes information about one method for predicting rival university Indicator and Final Scores, with the method selected via a drop-down list. Once a method is selected, the chart shows the predicted values, with the violin plot showing the uncertainty of that prediction as a distribution across all possible scores for that Indicator. This allows the user to  compare the rival universities expected value with the range of predictions made in the selected scenario. We can select a particular rival to have their values highlighted in the radar chart, for ease of use.

Having two views allows the user to go from a higher-level view of all methods across all rivals in the heat map, to a more focused view of one method in the radar chart. This abides by the "detail on demand" principle, as the user only goes to the radar char once they have noticed a method they want to focus on in the heat map.

\subsection{Implementation Details}
Our visual system was built using Vue.js, with a Python backend server. All the models were built using Python with scikit-learn.

\section{Evaluation}

To evaluate our system, we conducted a case study and an interview with the experts to evaluate the overall effectiveness of the system.
\begin{figure}
 \centering
 \includegraphics[width=\linewidth]{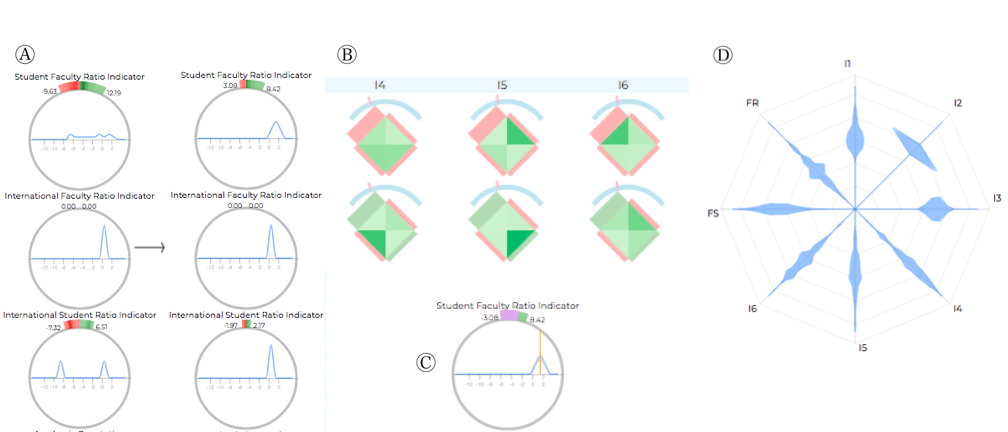}
 \caption{The main findings in the case study. (A) shows the variance in prediction for SFRI, and the two peaks in ISRI. It also shows the change in distribution after the filtering is applied. (B) shows the differences in the two selected scenarios for I4, I5, and I6. The highlighted quadrant corresponds to the first attribute. (C) shows the selected scenarios' uncertainty bound and position relative to other scenarios. (D) shows the distribution of predictions for the selected scenario, with the focus being on I1's shape.}
 \label{fig:case_studies}
 \end{figure}
\subsection{Finding Optimal Data Submission For a Ranking System}

 University A (``Uni") was in the process of submitting data to QS. They have a range of values that could be submitted for each attribute, and wanted to use the system to decide what to submit. The user was a female expert from Uni. 

She first selected Uni and inputted her submission range in the panel at the far left of Fig.\ref{fig:teaser}, which generated 1511 scenarios \textbf{(R2)}.

She noticed that the expected changes in Student Faculty Ratio Indicator (``SFRI") and International Student Ratio Indicator (``ISRI") had a lot of variation, while the other indicators didn't \textbf{(R2)}. She therefore decided to keep only those scenarios that have a predictive value larger than 0 for those two indicators, using the Scenario Analysis View.  She then noticed that there are two peaks of predicted values for ISRI, so she focused in on the second peak \textbf{(R2)}. The filtering had left 108 scenarios Fig.\ref{fig:case_studies}(A). She expressed surprise that there was such a large spread of values for SFRI and ISRI, stating this was an interesting insight. 

She now focused on the differences between the scenarios that remained after filtering. She decided to focus on the influence of the first attribute, as Uni had the most leeway on this attribute, so she sorted by this attribute and selected the first and last scenario to analyse more closely \textbf{(R2)} Fig.\ref{fig:case_studies}(B). She immediately noticed that the Local Influence was different in I4, I5, and I6.The change in the attribute had also caused the influence of other attributes to change, with some increasing and others decreasing Fig.\ref{fig:case_studies}(B). In particular, when the chosen attribute was the main influencer of I6 only when it had a negative change \textbf{(R1)}. She noted it down for further analysis with her team later. 

She now wanted to focus on the values for submission, so she selected the scenario matching one of the discrete values, to view it in the Scenario Analysis View in context of all scenarios \textbf{(R2)}Fig.\ref{fig:case_studies}(C). She immediately noticed that for SFRI, her selected scenario seemed to be predicting the average value, but the uncertainty was heavily biased in the negative direction \textbf{(R1)}. This bias in the negative direction implies that there is a significant possibility of the real outcome being below the predicted value, and so it could be risky to completely rely on the predicted value. She further investigated the scenario using the radar chart Fig.\ref{fig:case_studies}(D). Here she noticed that for I1, there is a bottom heavy distribution, so in her view that was enough evidence to discount this scenario with regards to I1. Similarly, the other values are viewed and, based on analysis in the same vein as above, she decided on a particular value set for data submission.
 
The case study shows how insights can be gained with regards to the optimality of the data points in the set of submittable values, as well as with regards to the relationship between the attributes and the indicators. 

\section{Discussion and Conclusion}

In this paper, we propose RankBooster, an interactive visual analytics system to help users conveniently investigate ranking predictions to understand the reasons of the ranking. We take university rankings as an example, working closely with the domain experts to generate requirements and build out the system. Our case study demonstrates how RankBooster can be be used to analyze scenarios and decide on the optimal data submission.

However, it is not without limitations. First, our analysis assumes that the attributes are independent of each other. Whilst this assumption is used in our domain, it may not always hold in other domains. Second, our evaluation is limited to universities specifically, further evaluation would need to be done with other domains to fully verify its usability in a more general setting. Tackling these issues will form the bulk of our future work in this area.

\bibliographystyle{eg-alpha-doi}
\bibliography{egbibsample}


\end{document}